\documentclass[aps,prm,reprint,longbibliography]{revtex4-1}

\usepackage{amsmath}
\usepackage{graphicx} 
\usepackage[sort&compress]{natbib}
\usepackage{colortbl}
\definecolor{lightgray}{gray}{0.9}
\usepackage{xcolor}
\usepackage[%
        colorlinks=true,urlcolor=blue,citecolor=blue,
        pdfpagelabels=true,hypertexnames=true,
        plainpages=false,naturalnames=true
        ]{hyperref}
\usepackage{pdfpages}
\makeatletter
\AtBeginDocument{\let\LS@rot\@undefined}
\makeatother

\begin{document}


\title{Phonon-assisted diffusion in the bcc phase of titanium and zirconium from first-principles}


\author{Sara Kadkhodaei}
\email[]{To whom correspondence should be addressed; Email: sarakad@uic.edu}
\author{Ali Davariashtiyani}
\affiliation{University of Illinois at Chicago, Chicago, Illinois, USA}
\date{\today}
\begin{abstract}
Diffusion is the underlying mechanism for many complicated materials phenomena, and understanding it is basic to discovery of novel materials with desired physical and mechanical properties. Certain groups of solid phases, such as the bcc phase of IIIB and IVB metals and their alloys, which are only stable when they reach high enough temperatures and experience anharmonic vibration entropic effects, exhibit ``anomalously fast diffusion''. However, the underlying reason for the observed extraordinary fast diffusion is poorly understood and due to the existence of harmonic vibration instabilities in these phases the standard models fail to predict their diffusivity. Here, we indicate that the anharmonic phonon-phonon coupling effects can accurately describe the anomalously large macroscopic diffusion coefficients in the bcc phase of IVB metals, and therefore yield new understanding on the underlying mechanism for diffusion in these phases. We utilize temperature-dependent phonon analysis by combining \textit{ab initio} molecular dynamics with lattice dynamics calculations to provide a new approach to use the transition state theory beyond the harmonic approximation. We validate the diffusivity predictions for the bcc phase of titanium and zirconium with available experimental measurements, while we show that predictions based on harmonic transition state theory severely underestimates diffusivity in these phases.  
\end{abstract}

\maketitle

\section{Introduction}
Diffusion processes play a key role in the kinetics of many materials-related phenomena, such as micro-structural evolution during materials processing, and discovery of materials with specific diffusion characteristics is a major effort in various research communities. Understanding mechanisms of diffusion and providing an \textit{a priori} predictive ability for diffusivity are essential for discovery of materials with specific diffusion characteristics and first-principles methods have become a common approach to predict diffusivity in various materials~\cite{Mantina2008,Mantina2009,Wu2011,Wimmer2008,KOUDRIACHOVA2003,Domain2005,TUCKER2010,Sarnthein1996,CHOUDHURY2011,Blochl1990,VanderVen2008,VanderVen2001,Mendelev2009,Milman1993,Frank1996}. Existing computational frameworks based on the harmonic transition state theory~\cite{eyring1935,Vineyard1957} rely on the assumption that the solid phase is metastable and there exists no mechanical instabilities in the system. This assumption impedes the use of standard models to predict diffusivity in many solid-state systems~\cite{Grimvall2012,Einarsdotter1997,Mirgorodsky1997,Parlinski1997,Kadkhodaei2017,Persson2000,Craievich1997,Ye1987,Xu2007,Mirgorodsky1997,Quijano2009,KADKHODAEI2018,Zayak2003}, including the $\beta$ phase of group IIIB and IVB metals and their alloys, where the system exhibits harmonic phonon instabilities yet is stabilized at high enough temperatures due to strongly anharmonic lattice vibration entropic effects. Moreover, $\beta$ phase of IIIB and IVB metals and their alloys exhibit ``anomalously fast diffusion'', several orders of magnitude higher than fcc metals and other bcc metals, at temperatures far below their melting point~\cite{Vogl1984,Vogl1989,PEART1962,petry1990,herzig19872,Petry1988,Herzig1999}, yet the underlying reason for this anomaly is poorly understood.

A number of studies have proposed explanations for the anomalous fast diffusion in the bcc phase of IIIB and IVB metals~\cite{Gilder1975,Sanchez1975,SANCHEZ1978,Herzig1987,Vogl1990,Herzig1989,DUDAREV2008,Sangiovanni2019,Mendelev2010}. While some earlier studies proposed mixed vacancy mechanisms similar to contribution of divacancies jump or mobile defects in fcc metals as the basis of anomalous fast diffusion, later experimental evidence based on isotope effect measurements and neutron scattering in $\beta$ IIIB and IVB metals confirmed that the monovacancy mechanism is the predominant diffusion mechanism~\cite{Manke1982,Jackson1977,PEART1962,Vogl1989,petry1990,Petry1988}. Recently, Sangiovanni et al.~\cite{Sangiovanni2019} proposed a highly concerted stringlike atomic motion as the mechanism underlying anomalously large self-diffusivities in bcc Ti based on observations in an \textit{ab initio} molecular dynamics simulation. Other studies correlated the markedly fast diffusion to temperature-induced effects. For example, Sanchez and de Fontaine proposed a model which correlates diffusivity in $\beta$-Zr to the formation of the metastable $\omega$ phase (i.e., heterophase fluctuations between bcc and $\omega$ induced by anharmonic vibrations), in which the diffusion activation energy is assumed to be the formation free energy of $\omega$ embryo~\cite{Sanchez1975,SANCHEZ1978}. Herzig and co-workers concluded that the observed anomalous fast self-diffusion corresponds to the softening of the LA $\frac{2}{3}\left<111\right>$ phonon mode~\cite{Herzig1987,Herzig1989,Vogl1990}, supported by observation of higher diffusivity in phases with stronger softening effects~\cite{Herzig1987}. 

The contribution of this report is twofold; First we illustrate that while the monovacancy jump is the predominant mechanism for diffusion in the bcc phase of Ti and Zr, it is promoted by the anharmonic phonon-phonon coupling effects, which underlies the observed anomalously fast diffusion. Second, we provide a new approach based on first-principles calculations that can successfully predict the macroscopic diffusion coefficient for the bcc phase of titanium and zirconium where the standard computational frameworks based on harmonic transition state theory fail. We have previously shown that for bcc IVB metals and their compounds (which are mechanically unstable), temperature-induced dynamical stabilization results in hopping of the system among local distortions of the lattice (i.e., local minima) in a way that the average atomic positions stay at bcc~\cite{Kadkhodaei2017,KADKHODAEI2018}. These local minima are located along the trigonal path associated with the longitudinal L$\frac{2}{3}\left<111\right>$ phonon eigenvector~\cite{Kadkhodaei2017,KADKHODAEI2018}. Here we show that the coincidence of this dynamical hopping with the direction of vacancy jump along $\frac{1}{2}\left<111\right>$ nearest neighbor in bcc metals significantly promotes the macroscopic diffusivity. Our \textit{a priori} prediction is validated with available experimental diffusivity measurements. On the other hand, we indicate that predicting diffusivity based on harmonic lattice vibrations severely under-predicts this diffusivity. This further illustrates the significant role of anharmonic lattice vibration interactions in enhancing the diffusivity. Our findings shed light on understanding the fundamental origins of diffusion mechanism in entropy-stabilized phases. 
\section{Anharmonic lattice vibration effects within transition state theory}
The macroscopic diffusivity $D$, for a vacancy-mediated diffusion can be described in terms of the atomic jump distance and jump frequency~\cite{shewmon,MehrerBook}, as the following,
\begin{equation}\label{microDiff}
D=\alpha C_v a^2\Gamma
\end{equation}
where $\alpha$ is a geometrical factor ($\alpha=8/6$ for bcc), $a$ is the vacancy jump distance, $C_v$ is the equilibrium vacancy concentration, and $\Gamma$ is the successful vacancy jump frequency~\cite{shewmon}. Within the transition state theory~\cite{eyring1935,Vineyard1957,kadkhodaei2019-2,Mantina2008}, the defect jump frequency is expressed as $\Gamma=\nu^*\exp(-\Delta H_m/k_BT)$,
where $\Delta H_m$ is the migration enthalpy. $k_B$ is the Boltzmann's constant and $T$ denotes temperature. ${\nu}^*$ is the effective prefactor frequency, which encompasses all vibrational effects in the diffusion process. Within the harmonic transition state theory (HTST), ${\nu}^*$ is expressed as the ratio of the product of normal vibration frequencies (i.e., harmonic phonon frequencies) of the initial state to that of the non-imaginary normal frequencies of the transition state, $\nu^*=\frac{\prod_{i=1}^{3N}\nu_i}{\prod_{i=1}^{3N-1}\nu^{\prime}_i}$, 
where $\nu$ and $\nu^{\prime}$ are the frequencies for the initial and transition configurations, respectively, for a system of $N$ atoms.
The equilibrium monovacancy concentration, $C_v$, is calculated according to $C_v= \exp(-\frac{\Delta H_v-T\Delta S_v}{k_BT})$, where $\Delta H_v$ and $\Delta S_v$ are enthalpy and entropy of monovacancy formation, respectively. The formation enthalpy and entropy are calculated according to $\Delta H_v = H^{tot}(N-1)-\frac{N-1}{N}H^{tot}(N)$ and $\Delta S_v = S^{tot}(N-1)-\frac{N-1}{N}S^{tot}(N)$, respectively. $H^{tot}(N-1)$ and $S^{tot}(N-1)$ are the total enthalpy and entropy for a system with a monovacancy and $H^{tot}(N)$ and $S^{tot}(N)$ are the total enthalpy and entropy for the bulk phase, respectively. 
The vibration entropy for each system is obtained from $S_{vib}= -k_B\int_0^{\infty}\ln\left[2\sinh(\frac{\hbar\omega}{2k_BT})\right]g(\omega)d\omega$, where $\omega$ is the vibration frequency and $g(\omega)$ is phonon density of state (DOS). \\
\begin{figure*}[t]
 \includegraphics[width=\textwidth,height=0.8\textheight,keepaspectratio]{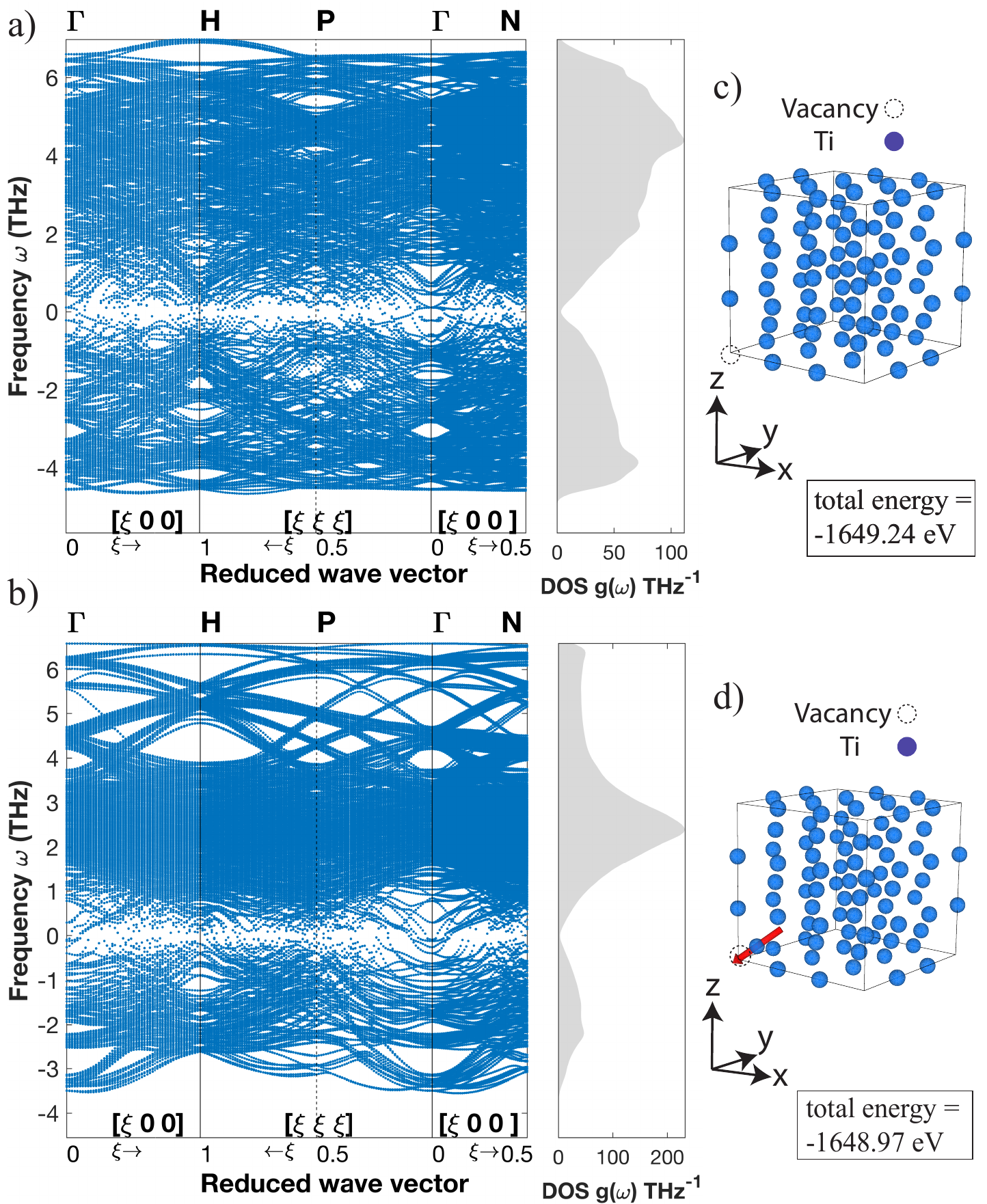}
 \caption{DFT harmonic phonon dispersion and density of states (DOS) for a $6\times6\times6$ supercell of bcc titanium including 215 atoms and a monovacancy a) at the bcc site, and b) migrated half way to the nearest neighbor bcc site along [111] direction. Phonon instabilities are depicted as negative frequencies in the dispersion curves. The atomic configuration for the initial and intermediate states of vacancy diffusion are illustrated in c) and d), respectively. A conventional supercell is presented for visualization simplicity.}
 \label{0kresults}
 \end{figure*}
 
 For systems that exhibit harmonic phonon instabilities, such as the bcc phase of IVB metals, calculation of diffusivity according to harmonic phonon analysis becomes infeasible. More specifically, the existence of multiple imaginary $\nu$ and $\nu^{\prime}$ frequencies impedes the use of HTST to obtain ${\nu}^*$, and $S_{vib}$ cannot be obtained based on imaginary phonon frequencies. 
In Fig.~\ref{0kresults}, the harmonic phonon dispersions for bcc Ti with a monovacancy at the initial and an intermediate state are obtained based on the density-functional theory (DFT) finite displacement approach, where the harmonic phonon dispersions exhibit phonon instabilities (similar plot for bcc Zr is presented in Fig. S5 in~\cite{Supplementray}). 
However, similar to other bcc IIIB and IVB metals, bcc Ti is stabilized at high enough temperature due to the entropic stabilization of the system arising from large amplitude anharmonic hopping among local distortions ~\cite{Kadkhodaei2017,KADKHODAEI2018,kadkhodaei2019-1,Ozolins2009,Einarsdotter1997,Mirgorodsky1997,Parlinski1997,Xu2007,Quijano2009,Grimvall2012,Craievich1997,Thomas2014,Hennig2008,Mei2009,Zhou2014,Duff2015,Liu2005}. We utilize two different variations of the self-consistent harmonic approximation (SCHA) approach, namely the self-consistent \textit{ab initio} lattice dynamics (SCAILD)~\cite{Werthamer1970,Souvatzis2008,Souvatzis2009} and the temperature dependent effective potential (TDEP) method~\cite{Hellman2011,Hellman2013}, to obtain the temperature-dependent phonon dispersion and DOS for well-defined equilibrium states, namely bulk bcc Ti and the bcc phase with a monovacancy (see section SI in~\cite{Supplementray} for more details).  
%
Fig.~\ref{allTphonon} compares the harmonic phonon dispersion (i.e., the zero-temperature phonon dispersion) of bcc Ti including phonon instabilities with finite-temperature phonon dispersions from 1300 K to 1600 K with no imaginary phonon frequencies, as expected due to dynamical stabilization of the phase (similar plot for bcc Zr is presented in Fig. S6 in~\cite{Supplementray}).

As illustrated in Fig.~\ref{allTphonon}(b)-(d), there is a negative dip at $\xi=\frac{2}{3}$ on the $L\left<\xi \xi \xi\right>$ phonon branch (depicted as the dashed line in Fig.~\ref{allTphonon}(d)), which is indicative of the phonon softening effects. 
This phonon mode corresponds to the trigonal distortion in which two neighboring (111) planes move towards each other and results in the $\omega$ phase when they collapse (see Fig~\ref{allTphonon}(e)). This atomic distortion coincides with the vacancy jump direction for the bcc phase. Accordingly, we assume that the effective prefactor frequency, $\nu^*$(T), which is the effective vibration frequency along the transition path, is equal to the temperature-dependent frequency of the L$\frac{2}{3}\left<1 1 1\right>$ phonon mode for the bulk bcc phase. 
%
 \begin{figure*}[t]
 \includegraphics[width=\textwidth]{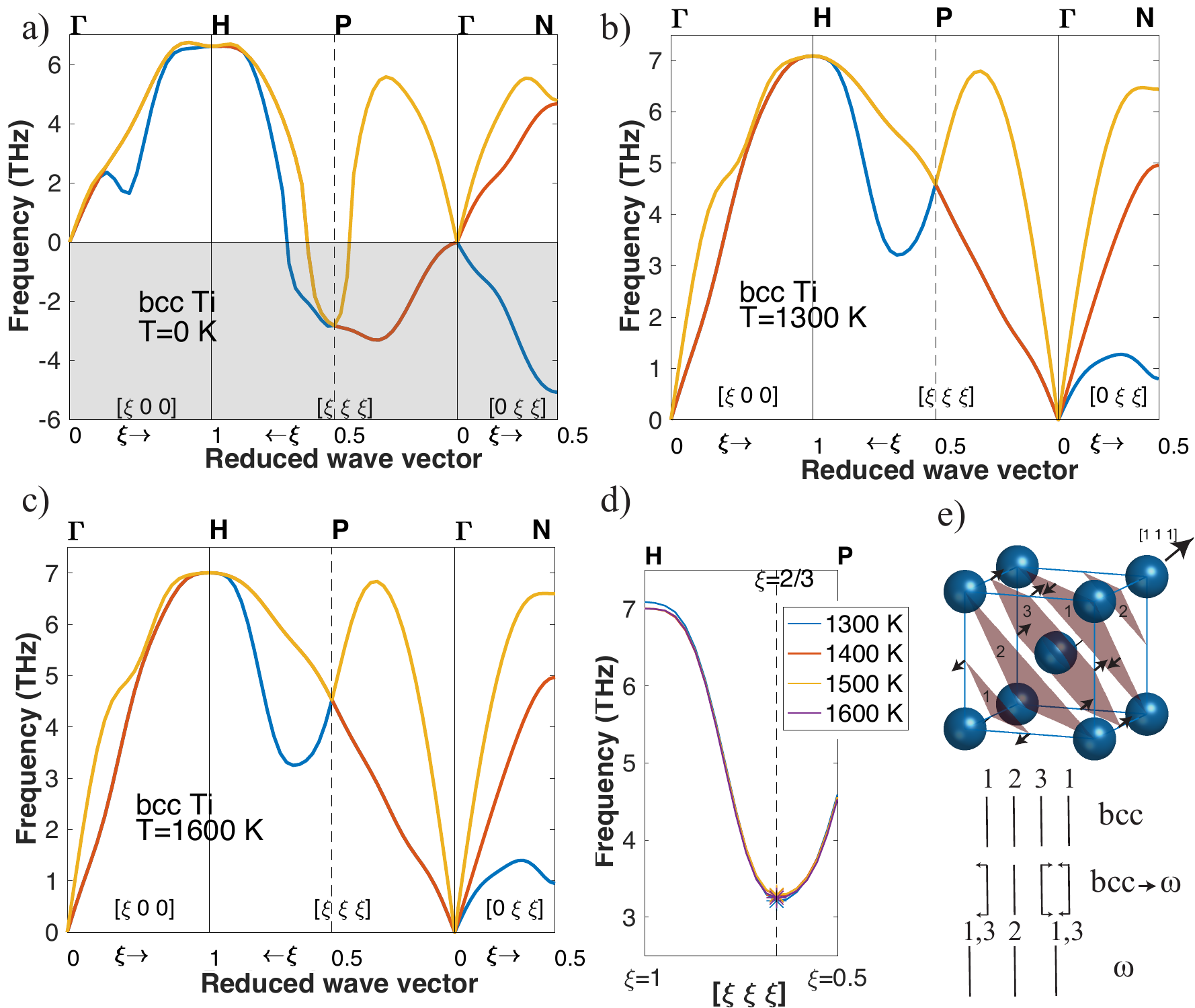}
 \caption{Phonon dispersion for a bulk defect-free $6\times6\times6$ supercell of bcc titanium including 216 atoms at a) 0 K, b) 1300 K, and c) 1600 K obtained using the SCAILD method. d) The longitudinal phonon branch along the $[\xi \xi \xi]$ reduced wave vector (between H and P high-symmetry points) from 1300 K to 1600 K. The star marks in (d) represent the negative dip on the phonon branch. e) The trigonal distortion of (111) atomic planes along the $[111]$ direction associates with the eigenvector of $\xi=\frac{2}{3}$ phonon mode along the $\Gamma$-P-H branch. A complete collapse of atomic planes 1 and 3 results in the hexagonal $\omega$ phase.}
 \label{allTphonon}
 \end{figure*}

The successful jump frequency, $\Gamma$, can then be obtained given the anharmonic effective frequency, $\nu^*$, and the activation free energy. To calculate the activation free energy in systems with harmonic phonon instabilities, there are some bottlenecks. Since the structure is mechanically unstable and only stabilizes at high temperatures, it cannot be relaxed according to zero-temperature forces to obtain the vacancy formation enthalpy. Additionally, common schemes such as the nudged elastic band (NEB) or dimer method~\cite{Henkelman2000,Henkelman2000a,Henkelman1999} cannot be used to locate the transition state as they rely on the assumption that the final and initial states of diffusion are metastable. Moreover, existence of imaginary phonon frequencies inhibits the calculation of vacancy formation entropy, $S_{vib}$. To overcome these problem, we use the statistical average energy and pressure from NVT \textit{ab initio} molecular dynamics (AIMD) simulations to obtain the enthalpy of the system at high temperature. In addition, vacancy formation entropy is obtained by comparing the temperature-dependent phonon frequencies and DOS for the bulk and defected systems. For the specific problem of bcc IVB metals, we assume that an interpolated state between the initial and final states (see Fig.~\ref{0kresults}(d)) is a reasonable approximation for the transition state. Accordingly, all the parameters needed for diffusivity calculation in Eq.~\ref{microDiff} can be obtained as temperature-dependent parameters that effectively account for phonon-phonon coupling effects: 
\begin{align}\label{finalD}
D=&\alpha a^2C_v \Gamma \nonumber \\
= &\alpha a^2 \nu^*(T)\exp(\frac{-\Delta H_m - \Delta H_v(T)}{k_BT})\exp(\frac{-\Delta S_v(T)}{k_B}).
\end{align}
 \section{Application to the bcc phase of IVB metals}\label{sec:app}
We applied the proposed methodology to obtain macroscopic diffusion coefficient for the bcc phase of titanium and zirconium. We calculate the migration enthalpy, $\Delta H_m$, as the DFT energy difference between the intermediate state and the initial state (represented in Fig.~\ref{0kresults}(c)-(d)). Note that we neglect the temperature-induced phonon-phonon coupling effects for the migration enthalpy calculation as the barrier energy for vacancy migration is calculated more accurately according to zero-temperature energies compared to the barrier energy for vacancy formation. This is because two more similar structures that both contain monovacancies are compared. Brute-force AIMD simulations cannot be used to calculate the enthalpy of the intermediate state since the simulation quickly slips away from this out-of-equilibrium state. A more sophisticated sampling technique is required which is the subject of a current development by the authors. The vacancy formation enthalpy is the difference of enthalpies for the bulk and defected systems and is obtained from \textit{ab initio} molecular dynamics (AIMD) average energy and pressure (see section SII in~\cite{Supplementray} for more details). Due to underestimation of DFT in obtaining the energy of the intrinsic surface formed around a vacancy, we add an explicit correction term to both the migration and vacancy formation enthalpies according to the electron density of the exchange-correlation functional~\cite{vac-corr,vac-corr-2} (see section SIII in~\cite{Supplementray} for more details). Fig.~\ref{Cv_compare}(a) illustrates the vacancy formation enthalpy for bcc titanium above the allotropic transformation temperature (a similar plot is presented for bcc zirconium in Fig. S9(d) in Supplemental Materials~\cite{Supplementray}).  
%
 
\begin{figure*}[t]
 \includegraphics[width=\textwidth,height=0.85\textheight,keepaspectratio]{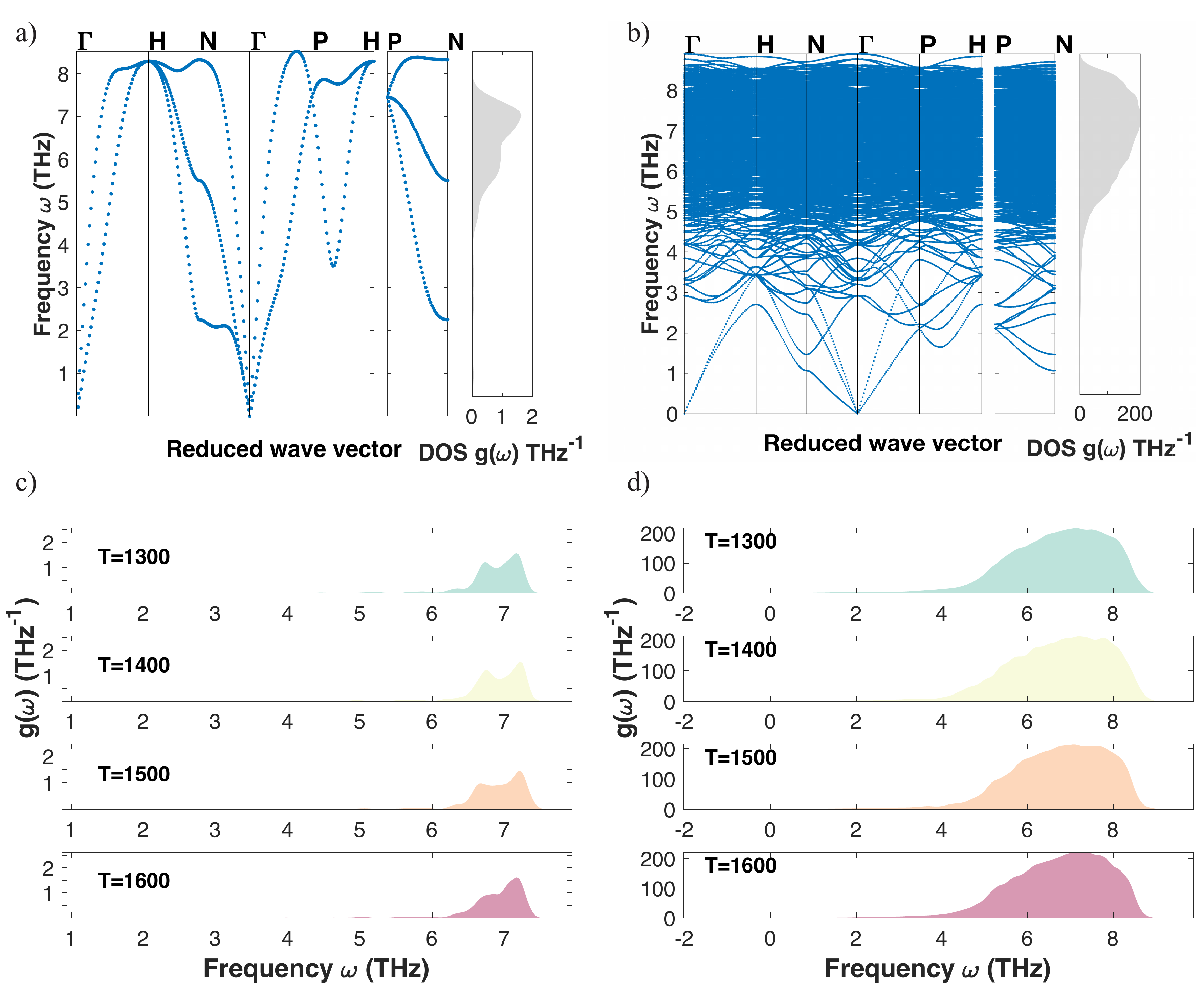}
 \caption{The anharmonic phonon dispersion and density of states (DOS) for a) a $6\times6\times6$ supercell of bcc titanium including 216 atoms and b) bcc Ti with a monovacancy including 215 atoms at 1600 K. c) The temperature-dependent DOS for bulk bcc Ti and d) the bcc phase with a monovacany at different temperatures.}
 \label{vacacnyformation}
 \end{figure*}

Vacancy formation entropy is obtained by comparing the temperature-dependent phonon frequencies and DOS for the bulk and defected systems. Fig.~\ref{vacacnyformation}(a)-(b) illustrates the temperature-dependent phonon dispersion and DOS for both systems at 1600 K obtained using the TDEP method~\cite{Hellman2011,Hellman2013} (similar plots for bcc Zr are shown in Fig. S7 in Supplemental Materials~\cite{Supplementray}). The formation of a vacancy results in appearance of optical phonon branches by removing the periodicity of the crystal structure and it broadens the peak in the DOS due to existence of several localized modes at higher frequencies. The vibration entropy for the bulk and defected systems are obtained based on T-dependent phonon frequencies, $\omega(T)$, and DOS, $g(\omega(T))$ (see section SIV in Supplemental Materials~\cite{Supplementray}). Fig.\ref{vacacnyformation}(c)-(d) shows the anharmonic phonon density of states for both the
bulk phase and the bcc phase including a monovacancy obtained by using the TDEP method at different temperatures (similar plots for bcc Zr are shown in Fig. S10 in Supplemental Materials~\cite{Supplementray}). Creating a vacancy in the supercell results in larger number of internal degrees of freedom, in this case $216\times3 - 3 = 645$, which is the area under the phonon density of states. The vacancy formation entropy for each temperature is shown in Fig.~\ref{Cv_compare}(b). As illustrated in Fig.~\ref{Cv_compare}(c), we calculate the equilibrium monovacancy concentration, $C_v$, having the formation entropy $\Delta S_v$ and formation enthalpy $\Delta H_v$ (see section SIV in Supplemental Materials~\cite{Supplementray}).  

  \begin{figure}[!h]
 \includegraphics[width=\columnwidth,height=0.85\textheight,keepaspectratio]{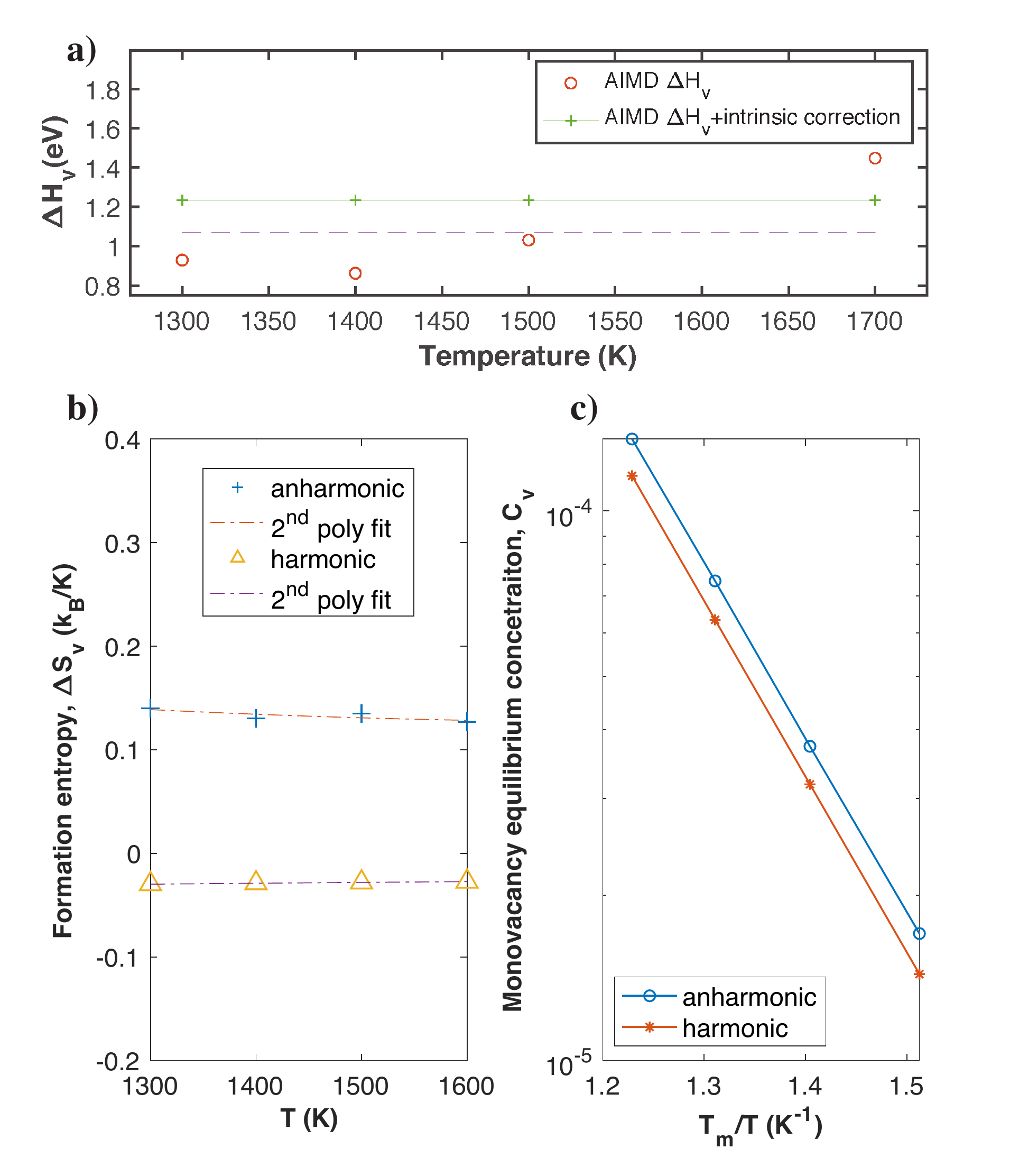}
 \caption{a) The vacancy formation enthalpy in bcc Ti as a function of temperature. The circles are the statical average formation enthalpy obtained from AIMD simulation using the PBE exchange-correlation functional. The crosses are the formation enthalpy values including the explicit intrinsic surface correction term. b) The formation entropy and c) equilibrium vacancy concentration versus temperature for monovacany in bcc Ti. For bulk bcc we used a $6\times6\times6$ supercell of bcc titanium including 216 atoms and for defected system we used bcc Ti with a monovacancy including 215 atoms.}
 \label{Cv_compare}
 \end{figure}

\begin{figure*}[t]
 \includegraphics[width=\textwidth,height=0.7\textheight,keepaspectratio]{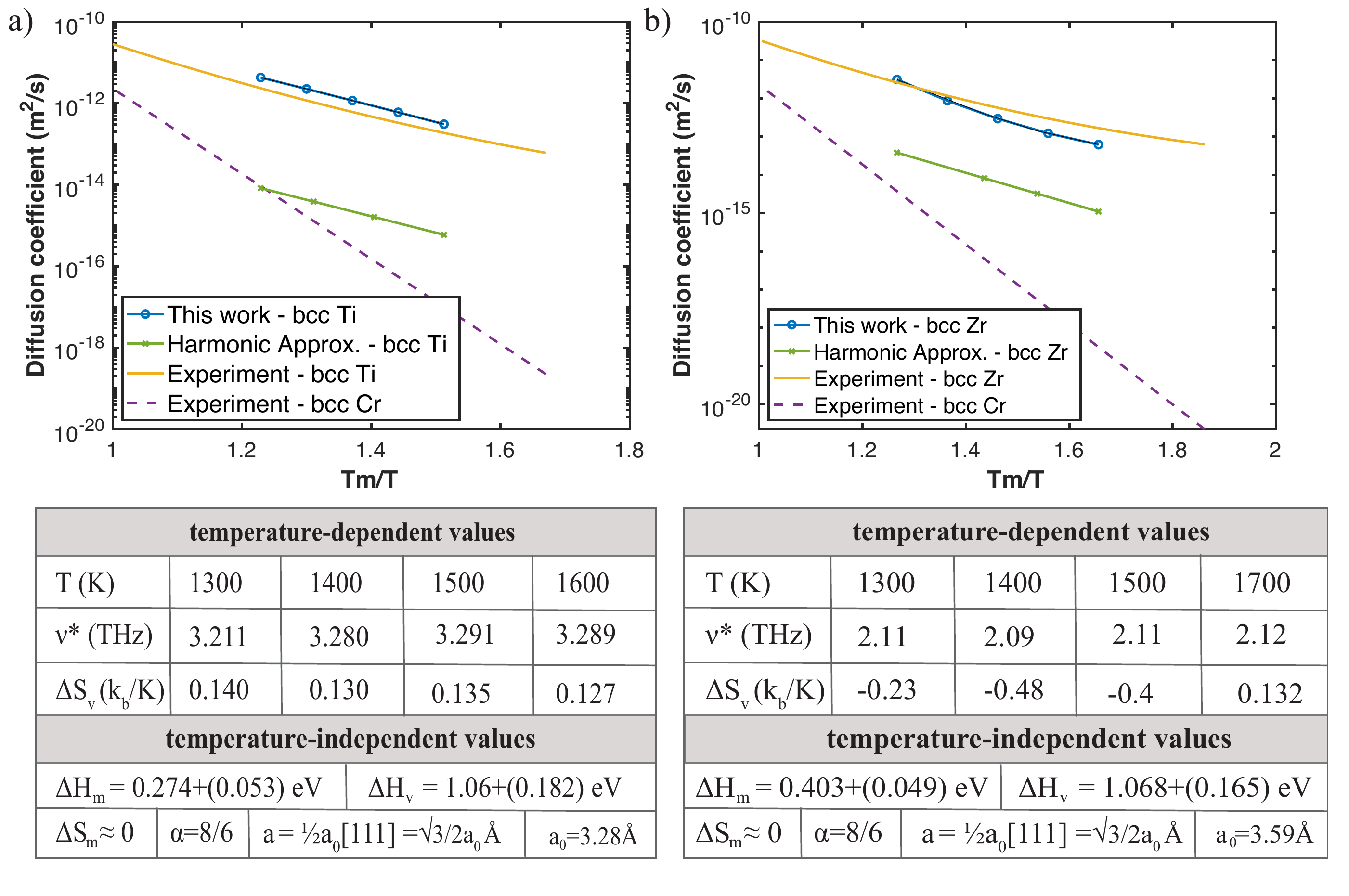}
  \caption{Complete DFT-based prediction of diffusion coefficient for a) bcc Ti and b) bcc Zr obtained according to anharmonic phonon analysis (circled-curves), validated with experimental measurements (solid curve) extracted from Ref.~\cite{Herzig1987}. Diffusion coefficient for bcc Cr (dashed curve) is also extracted from Ref.~\cite{Herzig1987} for comparison. Calculations based on standard harmonic phonon analysis (crosses) underestimates the diffusion coefficient by several orders of magnitude. $T_m$ is equal to 1966 K and 2153 K for Ti and Zr, respectively. All the parameters for calculating the diffusion coefficient are presented in the tables. The internal surface correction terms for vacancy formation and migration enthalpies are presented in parentheses. $\nu^*$ for each temperature is also represented by star marks in Fig.~\ref{allTphonon}(d) for bcc Ti and Fig. S6 for bcc Zr.}
 \label{diffusion}
 \end{figure*}

The \textit{ab initio} predicted diffusivity for bcc titanium and zirconium alongside the parameters used in Eq.~\ref{finalD} are presented in Fig.~\ref{diffusion}.
The circled-curve in Fig.~\ref{diffusion} are the diffusivity predictions by including the anharmonic lattice vibration effects, which agrees well with experimental measurements of Ref.~\cite{Herzig1987}. The diffusivity prediction based on the harmonic phonon analysis underestimates the diffusion coefficient by several orders of magnitude as shown by the crossed curve in Fig.~\ref{diffusion}. In spite of the fact that we use identical migration and formation enthalpies in both approaches, namely the proposed approach based on anharmonic phonon analysis and the standard harmonic model, the calculated diffusivities differ dramatically. Moreover, the vacancy formation entropy and concentration according to the proposed approach and the harmonic model are very similar for both bcc Ti and Zr (see Fig.~\ref{Cv_compare}(b)-(c) and Fig. S11 in Supplemental Materials~\cite{Supplementray}). This further illustrates the significant role of anharmonic lattice vibrations in enhancing diffusivity in the bcc phase of Ti and Zr. The harmonic lattice vibration model underestimates the diffusivity by almost 3 orders of magnitude and 2 orders of magnitude for bcc Ti and bcc Zr, respectively (see next section for more details on harmonic model prediction).
\section{Significance of phonon-phonon coupling effects}
To understand the significance of anharmonic vibration effects, we also calculated the diffusion coefficient according to a standard harmonic phonon analysis by excluding the imaginary harmonic phonon frequencies to calculate the vacancy formation entropy (see section SIV in~\cite{Supplementray}). We also use the T-independent value $\nu_1\times \frac{\prod_i\nu_i}{\prod_i\nu^{\prime}_i}$ to obtain the $\nu^*$ in Eq.~\ref{finalD} by assuming imaginary $\nu$ and $\nu^{\prime}$ frequencies have real values. $\nu_1$ is the most unstable phonon frequency (i.e., the square root of the most negative eigenvalue of the dynamical matrix) for the transition state. The transition state is assumed to be the estimated intermediate state presented in Fig.~\ref{0kresults}(d) for bcc Ti and in Fig. S5(d) for bcc Zr. We calculate $\nu_1$ to be 6.64 THz and 4.63 THz for Ti and Zr, respectively. This value is comparable to the temperature-dependent $\nu^*$ in our approach, presented in Fig.~\ref{diffusion}. However, the effective jump prefactor $\nu^*$ is severely low in the harmonic approximation. This is due to the small value for $\frac{\prod_i\nu_i}{\prod_i\nu^{\prime}_i}$, which essentially assumes phonon modes are independent (decoupled) oscillators and neglect phonon-phonon coupling effects. This assumption is severely underperforming for calculating diffusion coefficient in phases which harmonic phonon instabilities not simply due to infeasibility of its application but more importantly because of the dominant effect of anharmonic phonon interactions. We calculate the ratio of the product of independent harmonic phonon frequencies in the initial and intermediate states,$\frac{\prod_i\nu_i}{\prod_i\nu^{\prime}_i}$, to be 0.0011 and 0.0073 for bcc Ti and Zr, respectively, which results in $\nu^*$ values of 0.0073 THz and 0.034 THz for bcc Ti and Zr, respectively. Typical ratio values of $\frac{\prod_i\nu_i}{\prod_i\nu^{\prime}_i}$ for solid phases that are mechanically stable and can be accurately described by harmonic phonon modes are orders of magnitude higher. For example, the ratio value for fcc Al is 5.6 and for fcc Ag is 6.94. We obtained these values by comparing the DFT harmonic phonon eigenvalues for the initial and saddle point configuration obtained from the nudged elastic band method and are reported in our previous work~\cite{kadkhodaei2019-2}.
\section{Computational Details}
\subsection{DFT calculation}
Electronic structure calculations are performed using density functional theory (DFT) as implemented in the Vienna Ab-initio Simulation Package (VASP)~\cite{vasp1,vasp2,vasp3,vasp4}. We use the projector augmented wave (PAW) method~\cite{paw,paw_vasp} with energy cutoff of 274.6 eV for Ti and 229.9 eV for Zr and the generalized gradient approximation (GGA)~\cite{PERDEW1992,vasp3,vasp4} with the PBE~\cite{pbe} exchange-correlation (Ti\textunderscore{pv} and Zr\textunderscore{sv}). All DFT calculations, including static energy, molecular dynamics and SCAILD are performed on a $6\times6\times6$ supercell of bcc including 216 atoms and a supercell with a vacancy at the bcc site. For bcc Cr, we use the energy cutoff of 395.5 eV for the PAW psuedopotential with the PBE~\cite{pbe} exchange-correlation functional for a supercell of 128 atoms for the bulk phase and 127 atoms for the bcc phase with a monovacancy. 
\subsection{Enthalpy calculations}
AIMD simulations are performed at constant volume and temperature (NVT) under the Nose-Hoover thermostat~\cite{HOOVER1996,nose-1,nose-2,nose-3,nose-4}. Enthalpy values for the bulk and defected systems are obtained by averaging AIMD kinetic and potential energies and the pressure-volume term that are captured every 0.1 ps for a 35 ps trajectory after 1 ps thermalization at each temperature (see section SII in~\cite{Supplementray} for more details). 
\subsection{Phonon analysis}
SCAILD iteratively converges the phonon frequencies at different temperatures accounting for anharmonic effects~\cite{Werthamer1970,Souvatzis2008,Souvatzis2009}. In TDEP, temperature-dependent force-constant tensor is obtained by minimizing the difference between \textit{ab initio} molecular dynamics (AIMD) forces and forces described in the harmonic approximation~\cite{Hellman2011,Hellman2013} (see section SI in~\cite{Supplementray} for details).
SCAILD method is performed using the \texttt{SCPH} code~\cite{Souvatzis2008} based on DFT calculations until the free energy difference converges to $10^{-5}$ eV (see section SI in~\cite{Supplementray}). For the DFT calculations the Brillouin zone integration is performed on a mesh density of 4000 points per reciprocal angstrom (generated by the \texttt{ezvasp} module~\cite{atat1}). 
In TDEP method, phonon dispersions and DOSs are obtained from temperature-dependent force constant tensors using the \texttt{Phonopy} code~\cite{phonopy}. The temperature-dependent force constant tensor is obtained by least square fitting of AIMD forces and displacements to a harmonic model using our in-house code. For AIMD simulations, the first Brillouin zone integration is performed on the $\Gamma$ point for a $6\times6\times6$ supercell of bcc with cell volume of 3799.45 $\AA^3$ and 4988.63 $\AA^3$ for Ti and Zr, respectively.

 \section{Discussion}
The agreement of our first-principles prediction with experimental diffusion coefficient indicates that monovacancy jumps are the dominant atomic mechanism responsible for macroscopic diffusion in the bcc phase of IIIB-IVB metals. The anomalously larger diffusivity in bcc IIIB-IVB metals compared to other bcc and fcc metals can be accurately accounted for by including temperature-dependent anharmonic lattice vibration effects within the transition state theory to describe monvacancy jumps. In other words, monovacancy jumps are significantly promoted due to phonon-phonon coupling effects in these phases. On the other hand, the nature of free energy surface in bcc IIIB-IVB metals, which consists of multiple local minima around the bcc structure, results in hopping of the system among local structural distortions (this has been shown for bcc Ti and B2 NiTi in our previous studies~\cite{Kadkhodaei2017,KADKHODAEI2018}) and increases the likelihood of collective atomic motion. This phenomenon has been observed as heterophase fluctuations in earlier works~\cite{Vogl1990,Heiming1989} and as a concerted string-like atomic motion in a more recent \textit{ab initio} molecular dynamics simulation~\cite{Sangiovanni2019}. However, our calculations shows that these collective motions do not contribute to the macroscopic diffusivity, in other words they do not result in a long range mass transport. The contribution of collective atomic motion to diffusivity can be the subject of a future investigation which requires long-time molecular dynamics simulation (which are feasible based on potential models) for a complete statistical sampling of the diffusion process (similar to methods used to distinguish cyclic collective motion in tetragonal Li7La3Zr2O12 (LLZO)~\cite{PhysRevLett.116.135901}). One possibility is that these collective atomic motions are cyclic or has a random nature with a zero long-range effect. Unfortunately, the current potential models for bcc Ti (e.g., the MEAM potential~\cite{MEAM}) cannot effectively capture the softening of $L\frac{2}{3}\left<1 1 1\right>$ phonon, which plays the major role in softening of restoring forces and collective atomic distortions. The phonon-assisted vacancy jump mechanism to be the predominant one in bcc IVB metals is supported by direct experimental determination of the mechanism~\cite{PEART1962,Vogl1989,petry1990,Petry1988} and the isotope effect~\cite{Manke1982,Jackson1977}. 

While self-consistent harmonic approximation (SCHA) approaches (e.g., SCAILD and TDEP) have been extensively used for thermodynamic description of various systems~\cite{PhysRevB.94.020303,Hellman2011,Hellman2013,Souvatzis2009,Souvatzis2008,Errea2014,PhysRevB.96.014111,KOROTAEV201847}, we utilize self-consistent temperature-dependent phonon analysis to describe diffusive jump rates within the framework of the transition state theory for the first time to the best of our knowledge. The introduced approach can be used for diffusion description of other mechanically unstable phases, including the $\beta$ phase of IIIB and IVB metals and their alloys. The computation cost of the introduced methodology is minimal, only requiring anharmonic phonon analysis of two systems (a bulk phase and a small supercell including a vacancy) at each temperature, when compared to diffusion description based on brute-force molecular dynamic simulation in the order of nano seconds (refer to Ref.~\cite{PhysRevLett.116.135901} for details on adequate MD sampling of diffusive events). 

Developing a general framework that effectively accounts for temperature-induced effects in diffusion description of mechanically unstable but dynamically stabilized phases is beyond the scope of this report and is the subject of an active investigation by the authors. The major assumption in calculation of diffusivity in this report is the approximation of migration energy barrier and the effective jump frequency along the migration path (details are described in section~\ref{sec:app}). These assumption are justified for the pure bcc phase due to the coincidence of diffusive jump direction and a specific phonon mode with softening effects.  

In summary, we predict the diffusivity of the bcc Ti and Zr by simply including temperature-induced anharmonic phonon-phonon coupling effects within the monovacancy-mediated diffusive jump process. Our predictions are in good agreement with experimental measurements while disregarding the anharmonic vibration effects results in orders of magnitude underestimation in diffusivity. Our findings shed light on the underlying mechanism responsible for markedly larger diffusion coefficient in bcc IIIB and IVB metals and provide a computationally efficient and accurate approach for macroscopic diffusivity calculations beyond the harmonic lattice vibration approximation. 
\onecolumngrid
\section{Acknowledgments}
\small
We would like to thank Dr. John W. Lawson and Dr. Justin Haskins from NASA Ames Research Center for kindly sharing their code for T-dependent force-constant tensor calculation. Our in-house code is a modification of their code. Some of the calculations in this work were performed on the Extreme Science and Engineering Discovery Environment (XSEDE) resource through allocation TG-DMR190017~\cite{xsede}. 
\twocolumngrid
\nocite{Frenkel,vac-corr-3,vac-corr-4,vac-corr-5,vac-corr-6}
%

\onecolumngrid
\includepdf[pages=1-18,pagecommand={},width=\paperwidth]{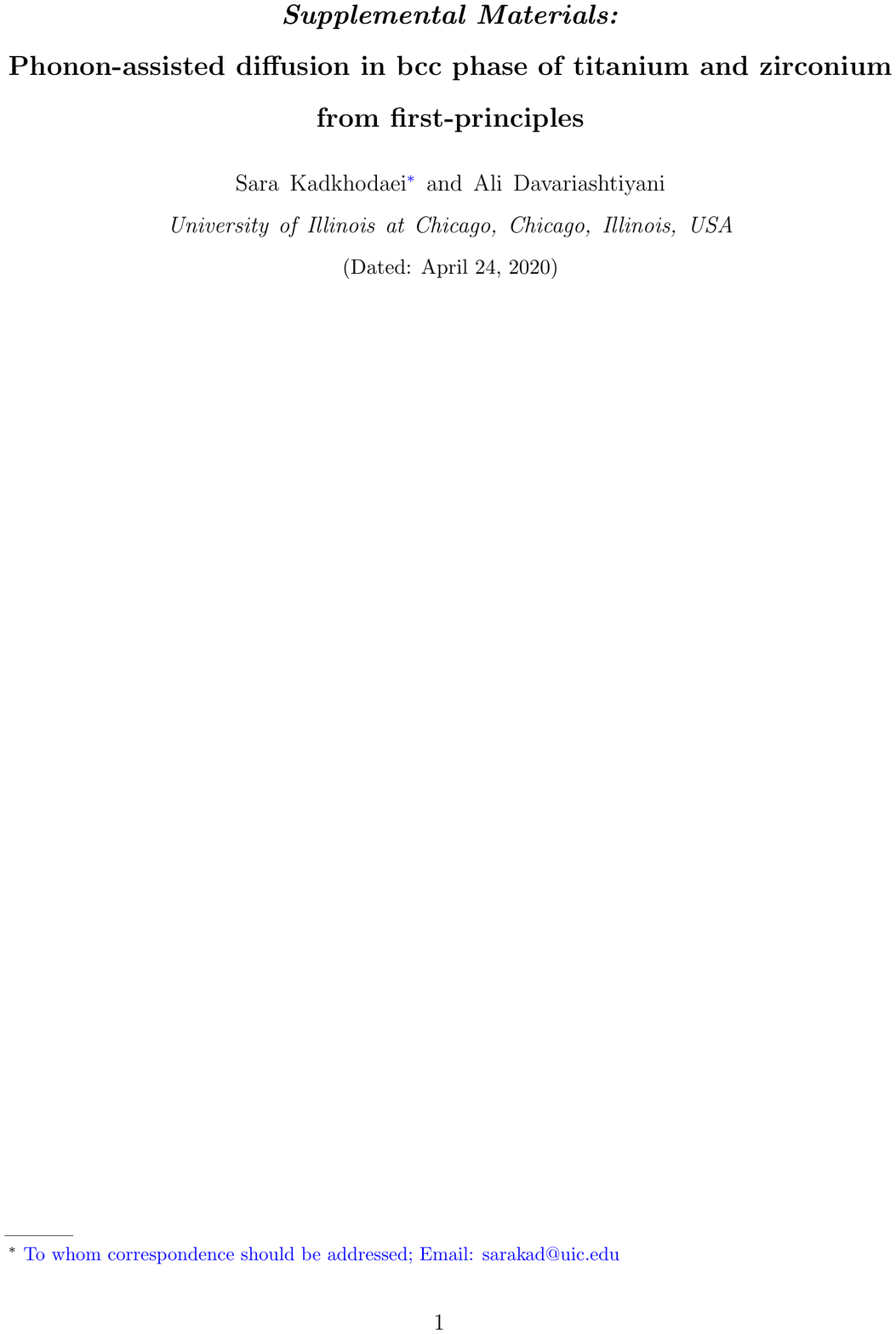}
\end{document}